\newcommand{\Tr}{\mathop{\mathrm{Tr}}\nolimits}

\documentclass[pra,twocolumn,showpacs,amsmath,amssymb,superscriptaddress]{revtex4-1}
\usepackage{graphicx,bm,color,mathptmx,hyperref}

\begin{document}

\title{Informational completeness of continuous-variable measurements}

\author{D.~Sych}
\affiliation{Max-Planck-Institut f\"ur die Physik des Lichts, 
G\"{u}nther-Scharowsky-Stra{\ss}e 1, Bau 24, 
91058 Erlangen, Germany}

\author{J.~\v{R}eh\'{a}\v{c}ek}
\affiliation{Department of Optics, Palack\'y University,
17. listopadu 12, 771 46 Olomouc, Czech Republic}

\author{Z.~Hradil}
\affiliation{Department of Optics, Palack\'y University,
17. listopadu 12, 771 46 Olomouc, Czech Republic}

\author{G.~Leuchs}
\affiliation{Max-Planck-Institut f\"ur die Physik des Lichts, 
G\"{u}nther-Scharowsky-Stra{\ss}e 1, Bau 24, 
91058 Erlangen, Germany}
\affiliation{Universit\"{a}t Erlangen-N\"{u}rnberg,
Staudtstra{\ss}e 7/B2, 91058 Erlangen, Germany}

\author{L.~L.~S\'{a}nchez-Soto} 
\affiliation{Max-Planck-Institut f\"ur die Physik des Lichts, 
G\"{u}nther-Scharowsky-Stra{\ss}e 1, Bau 24, 
91058 Erlangen, Germany}
\affiliation{Departamento de \'Optica, Facultad de F\'{\i}sica,
Universidad Complutense, 28040~Madrid, Spain}

\begin{abstract}
  We justify that homodyne tomography turns out to be informationally
  complete when the number of independent quadrature measurements is
  equal to the dimension of the density matrix in the Fock
  representation. Using this as our thread, we examine the
  completeness of other schemes, when continuous-variable observations
  are truncated to discrete finite-dimensional subspaces.
\end{abstract}


\maketitle

\textit{Introduction.---} 
Measurement lies at the very heart of quantum information.  It is an
indispensable tool for identifying how well one can prepare or create
a particular quantum state, as such an assessment is achieved by
performing suitable measurements on a sequence of identically prepared
systems. A set of measurements whose outcome probabilities are
sufficient to determine an arbitrary quantum state is called
informationally complete~(IC)~\cite{Prugovecki:1977fk,*Busch:1989kx},
while the process of reconstructing the state itself is broadly called
quantum tomography~\cite{lnp:2004uq}.

Investigations on IC measurements have been extensively carried out
for discrete Hilbert spaces~\cite{Ariano:2004kx,*Flammia:2005fk,
  *Weigert:2006zr,*Durt:2006ly}.  For continuous variables, the
archetypical example is homodyne tomography, first suggested in the
seminal paper of Vogel and Risken~\cite{Vogel:1989zr} and implemented
by Raymer and coworkers~\cite{Smithey:1993ly}.  Still, the
reconstruction of the measured state is inexorably accomplished in
some finite subspace chosen by an educated guess.

In many experimental situations one has some additional prior
information that can be efficiently used for reducing the number of
relevant unknown parameters, making the reconstruction much
simpler. For example, the knowledge about the Gaussian character of
the state reduces the problem to the evaluation of its $2 \times 2$
covariance matrix, which is formally analogous to the estimation of a
spin 1/2~\cite{Rehacek:2009ys}.  In the same vein, if the system can
be represented (say, in the Fock basis) by a finite density matrix,
the number of quadratures needed for an accurate reconstruction in
homodyne tomography is precisely equal to the dimension of that
matrix~\cite{Leonhardt:1996bh,*Leonhardt:1997cr}.  Note, in passing,
that this is indeed very closely related to the quantum version of the
sampling theorem~\cite{Oppenheim:2010ve}.

With the advent of powerful nonlinear estimation techniques, offering
amazing performance and robustness in most applications, matching the
signal space used for coding the information in an experimental setup
is more relevant than ever. This is, in fact, a dual problem: one is
not only interested in what can be reconstructed from a particular
scheme, but also in identifying the best option for a given signal.
Any measurement is represented by an operator and what matters for
tomography is the number of linearly independent operators required
for the reconstruction. Keeping in mind the lesson from homodyne
detection, the question of how to select those operators appears far
from trivial.

In this work, we examine the informational completeness of various
schemes. We derive a surprisingly simple connection between the
different measurement settings and the number of linearly independent
elements induced by them. In addition, the analysis displays the
information provided by each acquired piece. Such knowledge is
certainly of great interest both for the optimization of feasible
tomographic protocols and for the design of new ones.

\textit{Preliminaries.---}
A $d$-dimensional quantum system is represented by a positive
semidefinite $d \times d$ density matrix that requires $d^{2} -1$
independent real numbers for its specification.  A von Neumann
measurement fixes at most $d -1$ real parameters, so $d + 1$
tests have to be performed to reconstruct the state. This means
that $d^{2} + d$ histograms have to be recorded: the approach is thus
suboptimal for this number is higher than the number of parameters in
the density matrix. The von Neumann strategy can be further optimized
regarding this redundancy when the bases in which the measurements are
performed are mutually unbiased~\cite{Schwinger:1960dq,
*Wootters:1989qf,*Klimov:2007bh,*Klimov:2009ff,*Durt:2010cr}.

Besides, it is known that more general measurements exist.  Such
generalized measurements appeared as positive operator-valued measures
(POVMs) in the quantum theory of detection in the early 70s. It was
soon realized that, for numerous reasons, they provide greater
efficiency~\cite{Stratonovich:1973qe,Grishanin:1973hc,Helstrom:1976ij,
Holevo:2003fv,Braginski:1992nx}.

In short, a POVM is described by a set of linear operators $\{
\hat{E}_{n} \}$ furnishing the correct probabilities in any experiment
(we assume for simplicity discrete outcomes) through the Born rule
\begin{equation}
  p_{n}  = \Tr  ( \hat{\varrho} \, \hat{E}_{n}  ) \, ,
\end{equation}
for any state described by the density operator $\hat{\varrho}$. Here
$\Tr$ denotes the trace of a complex matrix. Compatibility with the
properties of ordinary probability imposes the requirements of
positivity, Hermiticity and resolution of the identity:
\begin{equation}
  \label{condPOVM}
  \hat{E}_{n} \ge 0 ,
  \qquad
  \hat{E}_{n} =  \hat{E}_{n}^\dagger ,
  \qquad
  \sum_{n} \hat{E}_{n} = \hat{\openone} \, .
\end{equation}
The case of projective von Neumann test is recovered when all the
operators in the set $\{ \hat{E}_{n} \}$ commute.

A quite general way of implementing a POVM is to properly combine the
system to be tested with an auxiliary system whose initial state is
known and then to perform a von Neumann test on the combined
system~\cite{Helstrom:1970fu}.  When the coupling with the auxiliar
system and the von Neumann measurement are judiciously chosen, one can
get IC POVMs, which in addition are optimal in the sense that minimize
the number of independent detections that must be obtained during the
tomographic process~\cite{Renes:2004nx,*Appleby:2011ys,*Salazar:2012tg}.  
Below, we shall exploit a similar idea to make IC measurements from
incomplete ones by cutting the dimensionality of the signal
space~\cite{Rehacek:2009ve}.

\textit{Quadrature measurement.---}
To get the backbone of our proposal, we start with states that can be
written as a finite sum in the Fock basis. We  wish to characterize a
$d$-dimensional subspace $\hat{\openone}_{d} =\sum_{n=0}^{d-1}
| n\rangle \langle n|$ from projections onto the quadrature eigenstates
$|x_\theta\rangle$, which can be expressed as
\begin{equation}
  \langle n|x_\theta\rangle \propto 
 H_{n} (x) e^{- x^2/2}e^{i n \theta} \, .
  \label{qmeas}
\end{equation}
Our task is find out how many different phase settings $\theta$ (also
referred to as cuts) are needed to make the scheme complete. It is
clear that the complex argument of $\langle n|x_\theta\rangle$ is
independent of $x$, so one quadrature can never generate a complete
POVM. For example, the three states $(|0\rangle\langle
0|+|1\rangle\langle 1|)/2$ and $(|0\rangle\pm i |1\rangle)/\sqrt{2}$
have the same position distribution and hence the position measurement
($\theta=0$) cannot discriminate among them. Obviously, since $\langle
n|x_{\theta=0}\rangle$ is real for all $x$, such projections fail to
generate the $\sigma_y$ component. This difficulty persists in higher
dimensions.

Fortunately enough, this topic can be fully analyzed in a closed
form. Indeed, the probability distribution of the homodyne detection
can be written, up to a normalization factor, as
\begin{equation}
  p (x,\theta)\propto \sum_{k,\ell = 0}^{d-1}
  \varrho_{k\ell} \, H_{k} (x) \, H_{\ell} (x) e^{-x^2} e^{i(k-\ell )\theta},
  \label{pxq}
\end{equation}
where $x$ is the measured amplitude. Let us take all matrix elements
$\varrho_{k\ell}$ non-zero and calculate the number $N$ of linearly
independent combinations of the density matrix (\ref{pxq}) that can be
generated from different choices of $x$ and $\theta$. For every
polynomial term with a maximum power $x^n$, we have the following
different combinations of Hermite polynomials [omitting for brevity
the associated factor $e^{i(k-\ell ) \theta}$]:
\begin{equation}
  \begin{array}{rllll}
    x^0: & H_{0}H_{0} \\
    x^1: & H_{0}H_{1},  \, H_{1}H_{0} \\
    x^2: & H_{0}H_{2},  \,  H_{2}H_{0}, \,  H_{1}H_{1} \\
    x^3: & H_{0}H_{3},  \,  H_{3}H_{0}, \,  H_{1}H_{2}, \, H_{2}H_{1}\\
    & \ldots \\
    x^{2d-4}: & H_{d-2}H_{d-2}, \, H_{d-1}H_{d-3}, \, H_{d-3}H_{d-1}\\
    x^{2d-3}: & H_{d-2}H_{d-1}, \, H_{d-1}H_{d-2}\\
    x^{2d-2}: & H_{d-1}H_{d-1}  .  \\
  \end{array}
\end{equation}

For a single quadrature $\theta_1$, the number $N_{1}$ is determined by the
term $H_{d-1}H_{d-1}$ with highest polynomial power and is equal to
$N_{1} =2d-1$ (the total number of first elements $H_{k}H_{l}$ in all
rows of the above array). Since we have {\em all} the polynomial
powers from 0 up to $2d-2$, there are no more linearly independent
elements. The second value $\theta_{2}$ gives us additionally
$N_{2}=2d-3$ linearly independent combinations of $\rho_{kl}$ (the
total number of second elements $H_{k}H_{l}$ in all rows), the third
value $\theta_3$ gives $N_3=2d-5$, and so on.

An alternative way to get the same result is by noticing that after
measuring the first quadrature, $2d-1$ linear combinations of
$\varrho_{k\ell}$ are fixed and there remain $d^2-1-(2d-1)=(d-1)^2-1$
free parameters entering the probability~(\ref{pxq}). This,
effectively, lowers the dimension of the problem by one. Thus $N_{2}
=2(d-1)-1 =2d-3$, $N_{3} =2(d-2)-1=2d-5$, etc. The procedure can be
repeated until the number of cuts $m$ equals $d$, when $N$ reaches its
maximum value of $d^2$.  One can draw an important conclusion from
this discussion: by measuring each new quadrature, the prior
information is updated in such a way that the dimension of the unknown
part of the system becomes one unity lesser.

In consequence, $m$ different phase settings induce $N=\sum_{k=1}^{m}
N_{k}$ linearly independent POVM elements; that is,
\begin{equation}
  N = \left\{ 
    \begin{array}{ll}
      m(2d-m) , & \quad m<d \, , \\
      & \\
      d^2 , & \quad m\geq d \, ,
    \end{array} 
\right.
  \label{Mdm}
\end{equation}
which can be conveniently summarized as in Table~\ref{POVMsize}. This
relation is the main achievement of this paper, as it provides a full
description of the IC for the problem at hand.

\begin{table}
  \caption{\label{POVMsize} The number of linearly independent POVM
    elements induced by $m$ different quadrature measurements in a
    $d$-dimensional Fock subspace. Bold font indicates IC POVMs.}
  \begin{ruledtabular}
    \begin{tabular}{ccccccc}
      $d$ &  $m=1$  & $m=2$ & $m=3$ & $m=4$ & $m=5$ & $m=6$ \\ \hline
      {2} & 3 &\textbf{4}&\textbf{4} &\textbf{4}&\textbf{4}&\textbf{4}\\
      {3} & 5 & 8 &\textbf{9}&\textbf{9}&\textbf{9}&\textbf{9}\\
      {4} & 7 &12&15&\textbf{16}&\textbf{16}&\textbf{16}\\
      {5} & 9 &16&21&24&\textbf{25}&\textbf{25}\\
      {6} &11&20&27&32&35&\textbf{36}\\
      {7} &13&24&33&40&45&48\\
      {8} &15&28&39&48&55&60
    \end{tabular}
  \end{ruledtabular}
\end{table}

As we can see, to fully characterize a $d$-dimensional Fock subspace
one needs $d$ different quadratures. For a fixed $m$, the number $N$
scales quadratically with $d$ until $d=m$, due to the completeness of
the generated POVMs.  For $d>m$ the size of POVM growth only linearly,
hence in the asymptotic case of sufficiently large dimensionality the
measurement becomes more and more incomplete.

In this respect, we wish to highlight that our results do not rely on
a particular reconstruction scheme, as in
Ref.~\cite{Leonhardt:1996bh,*Leonhardt:1997cr}, but rather they emerge
from the properties of the measurement itself and pervade all the
continuous-variable detection schemes.

\textit{Physical discussion.---}
The analysis presented so far has significant consequences, some of
which will be explored here. Notice that the IC quadrature operators
with $d$ phases approximate the structure of mutually unbiased
measurements. Without the Hilbert-space truncation, the projectors
with the same phase setting are orthogonal, while for different phases
they overlap. These complementary properties become modified in a
finite-dimensional subspace, although they approach their
continuous-variable counterpart as long as the cutoff $d$ is
sufficiently large.

Loosely speaking, the orthogonal measurements bring always a
completely new bit of information, whereas the overlapping ones verify
the previously acquired information and add a new piece thereof. This
is the essence of the tomography, where afterwards all the space is
covered by an ``operator'' mesh. Since the homodyne detection on
finite dimensional subspaces is feasible, this gives an excellent
opportunity to investigate the information gained in the process of
the measurement.

We emphasize that even a single quadrature actually generates IC
POVMs, provided that the signal space is properly matched to such a
detection. For example, Table~\ref{POVMsize} shows that in a
five-dimensional Fock subspace a single quadrature generates 9
independent POVM elements. These are IC in a suitably chosen
three-dimensional subspace. Put differently, a single quadrature
provides characterization of a qutrit in a five-dimensional space or
two entangled qubits in an nine-dimensional space. Similarly, all
Pauli matrices are induced by just two quadratures (say, $x$ and
$p$). Such kind of prior knowledge about the signal space can
significantly reduce the number of measurements required for the
complete state reconstruction. A particularly natural situation is the
case of symmetric/antisymmetric
subspaces~\cite{Sych:2009oq,*Sych:2009kl}: a complete characterization
of a two-qutrit state requires 9 different quadratures, whilst an
antisymmetric two-qutrit state can be inferred with just one
quadrature measurement.

The previous results can be simplified in some cases where prior
information is available.  For instance, some elements of the density
matrix can be known to vanish, and we can reach the complete POVM with
less settings.  Let us consider an analog of a squeezed vacuum of the
form
\begin{equation}
  \label{SqVac}
  | \Psi \rangle =c_{0}  | 0 \rangle  + c_{1}  | 4 \rangle + 
  c_{2} | 8 \rangle  + \ldots +  c_{k} | 4k \rangle  +\ldots \, .
\end{equation}
The coefficients $c_{k}$ gradually decrease with $k$, which makes
possible to set them to zero at some threshold value \mbox{$c_k=0$}
for $k \ge d$. The resulting state lives in a $d$-dimensional
subspace, so one may think that $m=d$ phase settings would be needed
to characterize that subspace. Surprisingly, the prior information
about photons being generated in quadruples modifies the structure of
the induced POVM and the number of linearly independent measurements
is found to be different from that calculated before. Let us have a
look at the case $d=3$. The coefficients in (\ref{pxq}) contain
polynomial powers up to $4(d-1) \times 4(d-1)=16$:
\begin{equation}
  \begin{array}{rl}
    x^0: & H_{0}H_{0} \\
    x^4: & H_{0}H_{4}, \, H_{4}H_{0} \\
    x^8: & H_{4}H_{4}, \, H_{0}H_{8} , \, H_{8}H_{0}\\
    x^{12}: & H_{4}H_{8} , \,  H_{8}H_{4}\\
    x^{16}: & H_{8}H_{8} . 
  \end{array}
\end{equation}
By writing explicitly the power expansion of the coefficients
$H_{0}H_{8}=256 x^8-3584 x^6+13440 x^4-13440 x^2+1680$ and
$H_{4}H_{4}= (16 x^4-48 x^2+12 )^2$, we see that they are linearly
independent. Indeed, $H_{0}H_{8}$ cannot be written as a linear
combination of $H_{0}H_{0}$, $H_{0}H_{4}$ and $H_{4}H_{4}$ due to the
lack of {\em all} needed polynomial powers, in contrast to the general
case. Hence, the number of linearly independent POVM elements is much
higher now: instead of $N_1=2d-1$, $N_2=2d-3$, $N_3=2d-5$, etc, we
have the modified expressions $N_{1}^\ast= N_{1} + N_{3} + N_{5}
+\ldots, N_{2}^\ast=N_{2}, N_{3}^\ast = N_{4}, N_{4}^\ast=N_{6},
\ldots$.  The number of different phase settings for the scheme to be
complete is $m^{\ast} = [d /2 ] + 1$, where $[ \, ]$ denotes the
integer part. Hence, the family of states (\ref{SqVac}) can be
characterized with roughly half as many quadratures that would be
required for a truncated Fock subspace of the same dimension.

As our last contention, we insist in that proper matching conditions
are essential~\cite{Ourjoumtsev:2006qa}. Particularly, without a
deeper understanding, too dense binning of quadratures may be useless,
whereas quadrature phase may easily become undersampled. In other
words, the number of quadrature bins multiplied by the number of
phases settings does not, in general, equal the number of free
parameters in the homodyne detection.

\textit{Photon-number resolving detection}.---
To obtain our main result (\ref{Mdm}) we have essentially relied on
the polynomial representation of the quadrature~(\ref{qmeas}), namely
that the probability distribution~(\ref{pxq}) can be written with help
of polynomials up to a certain power, defined by dimensionality of the
quantum state. This property holds also for many other
situations; indeed, almost any physical measurement satisfies this
property~\cite{Helstrom:1976ij}.  For example, an approximation of
coherent states $| \alpha \rangle = e^{-|\alpha |^2/2} \sum_{n=0}^{\infty}
\alpha^{n}/\sqrt{n!} | n \rangle $ by a finite sum, neglecting tiny
contributions of sufficiently high $n$, is essentially the same
situation as above: linearly independent POVM elements can be obtained
from projections of a finite-dimensional state onto a properly matched
basis.

In the simplest scenario, the strategy can be photon-number-resolving
detection. In fact, instead of (\ref{pxq}) we have now
\begin{equation}
\label{CohFock}
p ( \alpha, n ) = | \langle n | \alpha \rangle |^2 = 
e^{- |\alpha|^{2}} \, \frac{| \alpha |^{2n}}{n!} \, .
\end{equation} 
An obvious drawback of this choice is phase insensitivity, since 
coherent states with the same amplitude and different phases give
identical statistics. To avoid this degeneracy, the photon-number
detection can be preceded by a suitable
displacement~\cite{Wittmann:2010mi,*Wittmann:2010pi} or even more
complicated ``steering'' operation. A detailed examination of this
point is out of the scope of the present work.

\textit{Concluding remarks.---}  
We have addressed the fundamental question of how many independent
measurements can be generated by a given experimental setup. The
analysis clarifies the issue of necessary and sufficient complexity
for tomographical schemes. It turns out that the effective size of
quantum tomography for any finite-dimensional system can be related to
the number of different quadratures observed, which illustrates an
exciting link between discrete and continuous-variable systems.  We
stress that such a reconstruction cannot be accomplished without the
appropriate delimitation of the reconstruction space, since the space
touched by the observation exceeds then the amount of available data.

Our central result indicates that a single quadrature can be
discretized up to $2d -1$ bins for a $d$-dimensional systen, This
appears to be very similar to the classical Nyquist frequency in the
Kotelnikov-Shannon coding theorem. By finding a simple analytical rule
we have provided a full characterization of informational completeness
for a wide class of continuous-variable measurements and made a step
towards a better understanding of quantum resources in quantum optics
and quantum information processing.

The analysis is applicable for arbitrary dimensionality and thus
serves as a good alternative to other traditional approaches based,
e.g., on mutually unbiased bases and symmetric IC POVMs.

Finally, note that only data-acquisition issues have been addressed
here: the subsequent reconstruction must be capable to deal with
generic nonequivalent detections. The mathematical method best suited
for this purpose it the maximum likelihood estimation; however, even
in this case, the continuous data are discretized due to the very
nature of the measurement and the finite amount of memory and
computational time available. We believe that our theory can be
helpful in future optimizations of continuous-variable measurements.

Financial support from the EU FP7 (Grant Q-ESSENCE), the Spanish DGI
(Grant FIS2011-26786), the UCM-BSCH program (Grant GR-920992), the
Czech Technology Agency (Grant TE01020229), the Czech Ministry of
Trade and Industry (Grant FR-TI1/384) and the IGA of Palack\'y
University (Grant PRF-2012-005) is acknowledged.


\begin{thebibliography}{34}%
\makeatletter
\providecommand \@ifxundefined [1]{%
 \@ifx{#1\undefined}
}%
\providecommand \@ifnum [1]{%
 \ifnum #1\expandafter \@firstoftwo
 \else \expandafter \@secondoftwo
 \fi
}%
\providecommand \@ifx [1]{%
 \ifx #1\expandafter \@firstoftwo
 \else \expandafter \@secondoftwo
 \fi
}%
\providecommand \natexlab [1]{#1}%
\providecommand \enquote  [1]{``#1''}%
\providecommand \bibnamefont  [1]{#1}%
\providecommand \bibfnamefont [1]{#1}%
\providecommand \citenamefont [1]{#1}%
\providecommand \href@noop [0]{\@secondoftwo}%
\providecommand \href [0]{\begingroup \@sanitize@url \@href}%
\providecommand \@href[1]{\@@startlink{#1}\@@href}%
\providecommand \@@href[1]{\endgroup#1\@@endlink}%
\providecommand \@sanitize@url [0]{\catcode `\\12\catcode `\$12\catcode
  `\&12\catcode `\#12\catcode `\^12\catcode `\_12\catcode `\%12\relax}%
\providecommand \@@startlink[1]{}%
\providecommand \@@endlink[0]{}%
\providecommand \url  [0]{\begingroup\@sanitize@url \@url }%
\providecommand \@url [1]{\endgroup\@href {#1}{\urlprefix }}%
\providecommand \urlprefix  [0]{URL }%
\providecommand \Eprint [0]{\href }%
\providecommand \doibase [0]{http://dx.doi.org/}%
\providecommand \selectlanguage [0]{\@gobble}%
\providecommand \bibinfo  [0]{\@secondoftwo}%
\providecommand \bibfield  [0]{\@secondoftwo}%
\providecommand \translation [1]{[#1]}%
\providecommand \BibitemOpen [0]{}%
\providecommand \bibitemStop [0]{}%
\providecommand \bibitemNoStop [0]{.\EOS\space}%
\providecommand \EOS [0]{\spacefactor3000\relax}%
\providecommand \BibitemShut  [1]{\csname bibitem#1\endcsname}%
\let\auto@bib@innerbib\@empty
\bibitem [{\citenamefont {Prugove\v{c}ki}(1977)}]{Prugovecki:1977fk}%
  \BibitemOpen
  \bibfield  {author} {\bibinfo {author} {\bibfnamefont {E.}~\bibnamefont
  {Prugove\v{c}ki}},\ }\href@noop {} {\bibfield  {journal} {\bibinfo  {journal}
  {Int. J. Theor. Phys.}\ }\textbf {\bibinfo {volume} {16}},\ \bibinfo {pages}
  {321} (\bibinfo {year} {1977})}\BibitemShut {NoStop}%
\bibitem [{\citenamefont {Busch}\ and\ \citenamefont
  {Lahti}(1989)}]{Busch:1989kx}%
  \BibitemOpen
  \bibfield  {author} {\bibinfo {author} {\bibfnamefont {P.}~\bibnamefont
  {Busch}}\ and\ \bibinfo {author} {\bibfnamefont {P.~J.}\ \bibnamefont
  {Lahti}},\ }\href@noop {} {\bibfield  {journal} {\bibinfo  {journal} {Found.
  Phys.}\ }\textbf {\bibinfo {volume} {19}},\ \bibinfo {pages} {633} (\bibinfo
  {year} {1989})}\BibitemShut {NoStop}%
\bibitem [{\citenamefont {Paris}\ and\ \citenamefont
  {\v{R}eh\'a\v{c}ek}(2004)}]{lnp:2004uq}%
  \BibitemOpen
  \bibinfo {editor} {\bibfnamefont {M.~G.~A.}\ \bibnamefont {Paris}}\ and\
  \bibinfo {editor} {\bibfnamefont {J.}~\bibnamefont {\v{R}eh\'a\v{c}ek}},\
  eds.,\ \href@noop {} {\emph {\bibinfo {title} {Quantum State Estimation}}},\
  \bibinfo {series} {Lect. Not. Phys.}, Vol.\ \bibinfo {volume} {649}\
  (\bibinfo  {publisher} {Springer},\ \bibinfo {address} {Berlin},\ \bibinfo
  {year} {2004})\BibitemShut {NoStop}%
\bibitem [{\citenamefont {Ariano}\ \emph {et~al.}(2004)\citenamefont {Ariano},
  \citenamefont {Perinotti},\ and\ \citenamefont {Sacchi}}]{Ariano:2004kx}%
  \BibitemOpen
  \bibfield  {author} {\bibinfo {author} {\bibfnamefont {G.~M.~D.}\
  \bibnamefont {Ariano}}, \bibinfo {author} {\bibfnamefont {P.}~\bibnamefont
  {Perinotti}}, \ and\ \bibinfo {author} {\bibfnamefont {M.~F.}\ \bibnamefont
  {Sacchi}},\ }\href@noop {} {\bibfield  {journal} {\bibinfo  {journal} {J.
  Opt. B}\ }\textbf {\bibinfo {volume} {6}},\ \bibinfo {pages} {S487} (\bibinfo
  {year} {2004})}\BibitemShut {NoStop}%
\bibitem [{\citenamefont {Flammia}\ \emph {et~al.}(2005)\citenamefont
  {Flammia}, \citenamefont {Silberfarb},\ and\ \citenamefont
  {Caves}}]{Flammia:2005fk}%
  \BibitemOpen
  \bibfield  {author} {\bibinfo {author} {\bibfnamefont {S.~T.}\ \bibnamefont
  {Flammia}}, \bibinfo {author} {\bibfnamefont {A.}~\bibnamefont {Silberfarb}},
  \ and\ \bibinfo {author} {\bibfnamefont {C.~M.}\ \bibnamefont {Caves}},\
  }\href@noop {} {\bibfield  {journal} {\bibinfo  {journal} {Found. Phys.}\
  }\textbf {\bibinfo {volume} {35}},\ \bibinfo {pages} {1985} (\bibinfo {year}
  {2005})}\BibitemShut {NoStop}%
\bibitem [{\citenamefont {Weigert}(2006)}]{Weigert:2006zr}%
  \BibitemOpen
  \bibfield  {author} {\bibinfo {author} {\bibfnamefont {S.}~\bibnamefont
  {Weigert}},\ }\href@noop {} {\bibfield  {journal} {\bibinfo  {journal} {Int.
  J. Mod. Phys. B}\ }\textbf {\bibinfo {volume} {20}},\ \bibinfo {pages} {1942}
  (\bibinfo {year} {2006})}\BibitemShut {NoStop}%
\bibitem [{\citenamefont {Durt}(2006)}]{Durt:2006ly}%
  \BibitemOpen
  \bibfield  {author} {\bibinfo {author} {\bibfnamefont {T.}~\bibnamefont
  {Durt}},\ }\href@noop {} {\bibfield  {journal} {\bibinfo  {journal} {Open
  Sys. Inf. Dyn.}\ }\textbf {\bibinfo {volume} {13}},\ \bibinfo {pages} {403}
  (\bibinfo {year} {2006})}\BibitemShut {NoStop}%
\bibitem [{\citenamefont {Vogel}\ and\ \citenamefont
  {Risken}(1989)}]{Vogel:1989zr}%
  \BibitemOpen
  \bibfield  {author} {\bibinfo {author} {\bibfnamefont {K.}~\bibnamefont
  {Vogel}}\ and\ \bibinfo {author} {\bibfnamefont {H.}~\bibnamefont {Risken}},\
  }\href@noop {} {\bibfield  {journal} {\bibinfo  {journal} {Phys. Rev. A}\
  }\textbf {\bibinfo {volume} {40}},\ \bibinfo {pages} {2847} (\bibinfo {year}
  {1989})}\BibitemShut {NoStop}%
\bibitem [{\citenamefont {Smithey}\ \emph {et~al.}(1993)\citenamefont
  {Smithey}, \citenamefont {Beck}, \citenamefont {Raymer},\ and\ \citenamefont
  {Faridani}}]{Smithey:1993ly}%
  \BibitemOpen
  \bibfield  {author} {\bibinfo {author} {\bibfnamefont {D.~T.}\ \bibnamefont
  {Smithey}}, \bibinfo {author} {\bibfnamefont {M.}~\bibnamefont {Beck}},
  \bibinfo {author} {\bibfnamefont {M.~G.}\ \bibnamefont {Raymer}}, \ and\
  \bibinfo {author} {\bibfnamefont {A.}~\bibnamefont {Faridani}},\ }\href@noop
  {} {\bibfield  {journal} {\bibinfo  {journal} {Phys. Rev. Lett.}\ }\textbf
  {\bibinfo {volume} {70}},\ \bibinfo {pages} {1244} (\bibinfo {year}
  {1993})}\BibitemShut {NoStop}%
\bibitem [{\citenamefont {\v{R}eh\'a\v{c}ek}\ \emph {et~al.}(2009)\citenamefont
  {\v{R}eh\'a\v{c}ek}, \citenamefont {Olivares}, \citenamefont {Mogilevtsev},
  \citenamefont {Hradil}, \citenamefont {Paris}, \citenamefont {Fornaro},
  \citenamefont {D'Auria}, \citenamefont {Porzio},\ and\ \citenamefont
  {Solimeno}}]{Rehacek:2009ys}%
  \BibitemOpen
  \bibfield  {author} {\bibinfo {author} {\bibfnamefont {J.}~\bibnamefont
  {\v{R}eh\'a\v{c}ek}}, \bibinfo {author} {\bibfnamefont {S.}~\bibnamefont
  {Olivares}}, \bibinfo {author} {\bibfnamefont {D.}~\bibnamefont
  {Mogilevtsev}}, \bibinfo {author} {\bibfnamefont {Z.}~\bibnamefont {Hradil}},
  \bibinfo {author} {\bibfnamefont {M.~G.~A.}\ \bibnamefont {Paris}}, \bibinfo
  {author} {\bibfnamefont {S.}~\bibnamefont {Fornaro}}, \bibinfo {author}
  {\bibfnamefont {V.}~\bibnamefont {D'Auria}}, \bibinfo {author} {\bibfnamefont
  {A.}~\bibnamefont {Porzio}}, \ and\ \bibinfo {author} {\bibfnamefont
  {S.}~\bibnamefont {Solimeno}},\ }\href@noop {} {\bibfield  {journal}
  {\bibinfo  {journal} {Phys. Rev. A}\ }\textbf {\bibinfo {volume} {79}},\
  \bibinfo {pages} {032111} (\bibinfo {year} {2009})}\BibitemShut {NoStop}%
\bibitem [{\citenamefont {Leonhardt}\ and\ \citenamefont
  {Munroe}(1996)}]{Leonhardt:1996bh}%
  \BibitemOpen
  \bibfield  {author} {\bibinfo {author} {\bibfnamefont {U.}~\bibnamefont
  {Leonhardt}}\ and\ \bibinfo {author} {\bibfnamefont {M.}~\bibnamefont
  {Munroe}},\ }\href@noop {} {\bibfield  {journal} {\bibinfo  {journal} {Phys.
  Rev. A}\ }\textbf {\bibinfo {volume} {54}},\ \bibinfo {pages} {3682}
  (\bibinfo {year} {1996})}\BibitemShut {NoStop}%
\bibitem [{\citenamefont {Leonhardt}(1997)}]{Leonhardt:1997cr}%
  \BibitemOpen
  \bibfield  {author} {\bibinfo {author} {\bibfnamefont {U.}~\bibnamefont
  {Leonhardt}},\ }\href@noop {} {\bibfield  {journal} {\bibinfo  {journal} {J.
  Mod. Opt.}\ }\textbf {\bibinfo {volume} {44}},\ \bibinfo {pages} {2271}
  (\bibinfo {year} {1997})}\BibitemShut {NoStop}%
\bibitem [{\citenamefont {Oppenheim}\ \emph {et~al.}(2010)\citenamefont
  {Oppenheim}, \citenamefont {Willsky},\ and\ \citenamefont
  {Hamid}}]{Oppenheim:2010ve}%
  \BibitemOpen
  \bibfield  {author} {\bibinfo {author} {\bibfnamefont {A.~V.}\ \bibnamefont
  {Oppenheim}}, \bibinfo {author} {\bibfnamefont {A.~S.}\ \bibnamefont
  {Willsky}}, \ and\ \bibinfo {author} {\bibfnamefont {S.}~\bibnamefont
  {Hamid}},\ }\href@noop {} {\emph {\bibinfo {title} {Signals and Systems}}},\
  \bibinfo {edition} {2nd}\ ed.\ (\bibinfo  {publisher} {Prentice Hall},\
  \bibinfo {address} {New Jersey},\ \bibinfo {year} {2010})\BibitemShut
  {NoStop}%
\bibitem [{\citenamefont {Schwinger}(1960)}]{Schwinger:1960dq}%
  \BibitemOpen
  \bibfield  {author} {\bibinfo {author} {\bibfnamefont {J.}~\bibnamefont
  {Schwinger}},\ }\href@noop {} {\bibfield  {journal} {\bibinfo  {journal}
  {Proc. Natl. Acad. Sci. USA}\ }\textbf {\bibinfo {volume} {46}},\ \bibinfo
  {pages} {570} (\bibinfo {year} {1960})}\BibitemShut {NoStop}%
\bibitem [{\citenamefont {Wootters}\ and\ \citenamefont
  {Fields}(1989)}]{Wootters:1989qf}%
  \BibitemOpen
  \bibfield  {author} {\bibinfo {author} {\bibfnamefont {W.~K.}\ \bibnamefont
  {Wootters}}\ and\ \bibinfo {author} {\bibfnamefont {B.~D.}\ \bibnamefont
  {Fields}},\ }\href@noop {} {\bibfield  {journal} {\bibinfo  {journal} {Ann.
  Phys.}\ }\textbf {\bibinfo {volume} {191}},\ \bibinfo {pages} {363} (\bibinfo
  {year} {1989})}\BibitemShut {NoStop}%
\bibitem [{\citenamefont {Klimov}\ \emph {et~al.}(2007)\citenamefont {Klimov},
  \citenamefont {Romero}, \citenamefont {Bj{\"{o}}rk},\ and\ \citenamefont
  {S{\'{a}}nchez-Soto}}]{Klimov:2007bh}%
  \BibitemOpen
  \bibfield  {author} {\bibinfo {author} {\bibfnamefont {A.~B.}\ \bibnamefont
  {Klimov}}, \bibinfo {author} {\bibfnamefont {J.~L.}\ \bibnamefont {Romero}},
  \bibinfo {author} {\bibfnamefont {G.}~\bibnamefont {Bj{\"{o}}rk}}, \ and\
  \bibinfo {author} {\bibfnamefont {L.~L.}\ \bibnamefont
  {S{\'{a}}nchez-Soto}},\ }\href@noop {} {\bibfield  {journal} {\bibinfo
  {journal} {J. Phys. A}\ }\textbf {\bibinfo {volume} {40}},\ \bibinfo {pages}
  {3987} (\bibinfo {year} {2007})}\BibitemShut {NoStop}%
\bibitem [{\citenamefont {Klimov}\ \emph {et~al.}(2009)\citenamefont {Klimov},
  \citenamefont {Sych}, \citenamefont {S{\'a}nchez-Soto},\ and\ \citenamefont
  {Leuchs}}]{Klimov:2009ff}%
  \BibitemOpen
  \bibfield  {author} {\bibinfo {author} {\bibfnamefont {A.~B.}\ \bibnamefont
  {Klimov}}, \bibinfo {author} {\bibfnamefont {D.}~\bibnamefont {Sych}},
  \bibinfo {author} {\bibfnamefont {L.~L.}\ \bibnamefont {S{\'a}nchez-Soto}}, \
  and\ \bibinfo {author} {\bibfnamefont {G.}~\bibnamefont {Leuchs}},\
  }\href@noop {} {\bibfield  {journal} {\bibinfo  {journal} {Phys. Rev. A}\
  }\textbf {\bibinfo {volume} {79}},\ \bibinfo {pages} {052101} (\bibinfo
  {year} {2009})}\BibitemShut {NoStop}%
\bibitem [{\citenamefont {Durt}\ \emph {et~al.}(2010)\citenamefont {Durt},
  \citenamefont {Englert}, \citenamefont {Bengtsson},\ and\ \citenamefont
  {Zyczkowski}}]{Durt:2010cr}%
  \BibitemOpen
  \bibfield  {author} {\bibinfo {author} {\bibfnamefont {T.}~\bibnamefont
  {Durt}}, \bibinfo {author} {\bibfnamefont {B.-G.}\ \bibnamefont {Englert}},
  \bibinfo {author} {\bibfnamefont {I.}~\bibnamefont {Bengtsson}}, \ and\
  \bibinfo {author} {\bibfnamefont {K.}~\bibnamefont {Zyczkowski}},\
  }\href@noop {} {\bibfield  {journal} {\bibinfo  {journal} {Int. J. Quantum
  Inf.}\ }\textbf {\bibinfo {volume} {8}},\ \bibinfo {pages} {533} (\bibinfo
  {year} {2010})}\BibitemShut {NoStop}%
\bibitem [{\citenamefont {Stratonovich}(1973)}]{Stratonovich:1973qe}%
  \BibitemOpen
  \bibfield  {author} {\bibinfo {author} {\bibfnamefont {R.~L.}\ \bibnamefont
  {Stratonovich}},\ }\href@noop {} {\bibfield  {journal} {\bibinfo  {journal}
  {J. Stochastics}\ }\textbf {\bibinfo {volume} {1}},\ \bibinfo {pages} {87}
  (\bibinfo {year} {1973})}\BibitemShut {NoStop}%
\bibitem [{\citenamefont {Grishanin}(1973)}]{Grishanin:1973hc}%
  \BibitemOpen
  \bibfield  {author} {\bibinfo {author} {\bibfnamefont {B.~A.}\ \bibnamefont
  {Grishanin}},\ }\href@noop {} {\bibfield  {journal} {\bibinfo  {journal}
  {Tekn. Kibernetika}\ }\textbf {\bibinfo {volume} {11}},\ \bibinfo {pages}
  {127} (\bibinfo {year} {1973})}\BibitemShut {NoStop}%
\bibitem [{\citenamefont {Helstrom}(1976)}]{Helstrom:1976ij}%
  \BibitemOpen
  \bibfield  {author} {\bibinfo {author} {\bibfnamefont {C.~W.}\ \bibnamefont
  {Helstrom}},\ }\href@noop {} {\emph {\bibinfo {title} {Quantum Detection and
  Estimation Theory,}}}\ (\bibinfo  {publisher} {Academic},\ \bibinfo {address}
  {New York},\ \bibinfo {year} {1976})\BibitemShut {NoStop}%
\bibitem [{\citenamefont {Holevo}(2003)}]{Holevo:2003fv}%
  \BibitemOpen
  \bibfield  {author} {\bibinfo {author} {\bibfnamefont {A.~S.}\ \bibnamefont
  {Holevo}},\ }\href@noop {} {\emph {\bibinfo {title} {Probabilistic and
  Statistical Aspects of Quantum Theory}}},\ \bibinfo {edition} {2nd}\ ed.\
  (\bibinfo  {publisher} {North Holland},\ \bibinfo {address} {Amsterdam},\
  \bibinfo {year} {2003})\BibitemShut {NoStop}%
\bibitem [{\citenamefont {Braginski}\ and\ \citenamefont
  {Khalili}(1992)}]{Braginski:1992nx}%
  \BibitemOpen
  \bibfield  {author} {\bibinfo {author} {\bibfnamefont {V.~B.}\ \bibnamefont
  {Braginski}}\ and\ \bibinfo {author} {\bibfnamefont {F.~Y.}\ \bibnamefont
  {Khalili}},\ }\href@noop {} {\emph {\bibinfo {title} {Quantum Measurement}}}\
  (\bibinfo  {publisher} {Oxford University Press},\ \bibinfo {address}
  {Oxford},\ \bibinfo {year} {1992})\BibitemShut {NoStop}%
\bibitem [{\citenamefont {Helstrom}\ \emph {et~al.}(1970)\citenamefont
  {Helstrom}, \citenamefont {Liu},\ and\ \citenamefont
  {Gordon}}]{Helstrom:1970fu}%
  \BibitemOpen
  \bibfield  {author} {\bibinfo {author} {\bibfnamefont {C.~W.}\ \bibnamefont
  {Helstrom}}, \bibinfo {author} {\bibfnamefont {J.~W.~S.}\ \bibnamefont
  {Liu}}, \ and\ \bibinfo {author} {\bibfnamefont {J.~P.}\ \bibnamefont
  {Gordon}},\ }\href@noop {} {\bibfield  {journal} {\bibinfo  {journal} {Proc
  IEEE}\ }\textbf {\bibinfo {volume} {58}},\ \bibinfo {pages} {1578} (\bibinfo
  {year} {1970})}\BibitemShut {NoStop}%
\bibitem [{\citenamefont {Renes}\ \emph {et~al.}(2004)\citenamefont {Renes},
  \citenamefont {Blume-Kohout}, \citenamefont {Scott},\ and\ \citenamefont
  {Caves}}]{Renes:2004nx}%
  \BibitemOpen
  \bibfield  {author} {\bibinfo {author} {\bibfnamefont {J.~M.}\ \bibnamefont
  {Renes}}, \bibinfo {author} {\bibfnamefont {R.}~\bibnamefont {Blume-Kohout}},
  \bibinfo {author} {\bibfnamefont {A.~J.}\ \bibnamefont {Scott}}, \ and\
  \bibinfo {author} {\bibfnamefont {C.~M.}\ \bibnamefont {Caves}},\ }\href@noop
  {} {\bibfield  {journal} {\bibinfo  {journal} {J. Math. Phys.}\ }\textbf
  {\bibinfo {volume} {45}},\ \bibinfo {pages} {2171} (\bibinfo {year}
  {2004})}\BibitemShut {NoStop}%
\bibitem [{\citenamefont {Appleby}\ \emph {et~al.}(2011)\citenamefont
  {Appleby}, \citenamefont {Flammia},\ and\ \citenamefont
  {Fuchs}}]{Appleby:2011ys}%
  \BibitemOpen
  \bibfield  {author} {\bibinfo {author} {\bibfnamefont {D.~M.}\ \bibnamefont
  {Appleby}}, \bibinfo {author} {\bibfnamefont {S.~T.}\ \bibnamefont
  {Flammia}}, \ and\ \bibinfo {author} {\bibfnamefont {C.~A.}\ \bibnamefont
  {Fuchs}},\ }\href@noop {} {\bibfield  {journal} {\bibinfo  {journal} {J.
  Math. Phys.}\ }\textbf {\bibinfo {volume} {52}},\ \bibinfo {pages} {022202}
  (\bibinfo {year} {2011})}\BibitemShut {NoStop}%
\bibitem [{\citenamefont {Salazar}\ \emph {et~al.}(2012)\citenamefont
  {Salazar}, \citenamefont {Goyeneche}, \citenamefont {Delgado},\ and\
  \citenamefont {Saavedra}}]{Salazar:2012tg}%
  \BibitemOpen
  \bibfield  {author} {\bibinfo {author} {\bibfnamefont {R.}~\bibnamefont
  {Salazar}}, \bibinfo {author} {\bibfnamefont {D.}~\bibnamefont {Goyeneche}},
  \bibinfo {author} {\bibfnamefont {A.}~\bibnamefont {Delgado}}, \ and\
  \bibinfo {author} {\bibfnamefont {C.}~\bibnamefont {Saavedra}},\ }\href@noop
  {} {\bibfield  {journal} {\bibinfo  {journal} {Phys. Lett. A}\ }\textbf
  {\bibinfo {volume} {376}},\ \bibinfo {pages} {325} (\bibinfo {year}
  {2012})}\BibitemShut {NoStop}%
\bibitem [{\citenamefont {{\v R}eh{\'a}{\v c}ek}\ \emph
  {et~al.}(2009)\citenamefont {{\v R}eh{\'a}{\v c}ek}, \citenamefont {Hradil},
  \citenamefont {Bouchal}, \citenamefont {{\v C}elechovsk{\'y}}, \citenamefont
  {Rigas},\ and\ \citenamefont {S{\'a}nchez-Soto}}]{Rehacek:2009ve}%
  \BibitemOpen
  \bibfield  {author} {\bibinfo {author} {\bibfnamefont {J.}~\bibnamefont {{\v
  R}eh{\'a}{\v c}ek}}, \bibinfo {author} {\bibfnamefont {Z.}~\bibnamefont
  {Hradil}}, \bibinfo {author} {\bibfnamefont {Z.}~\bibnamefont {Bouchal}},
  \bibinfo {author} {\bibfnamefont {R.}~\bibnamefont {{\v C}elechovsk{\'y}}},
  \bibinfo {author} {\bibfnamefont {I.}~\bibnamefont {Rigas}}, \ and\ \bibinfo
  {author} {\bibfnamefont {L.~L.}\ \bibnamefont {S{\'a}nchez-Soto}},\
  }\href@noop {} {\bibfield  {journal} {\bibinfo  {journal} {Phys. Rev. Lett.}\
  }\textbf {\bibinfo {volume} {103}},\ \bibinfo {pages} {250402} (\bibinfo
  {year} {2009})}\BibitemShut {NoStop}%
\bibitem [{\citenamefont {Sych}\ and\ \citenamefont
  {Leuchs}(2009{\natexlab{a}})}]{Sych:2009oq}%
  \BibitemOpen
  \bibfield  {author} {\bibinfo {author} {\bibfnamefont {D.}~\bibnamefont
  {Sych}}\ and\ \bibinfo {author} {\bibfnamefont {G.}~\bibnamefont {Leuchs}},\
  }\href@noop {} {\bibfield  {journal} {\bibinfo  {journal} {New J. Phys.}\
  }\textbf {\bibinfo {volume} {11}},\ \bibinfo {pages} {013006} (\bibinfo
  {year} {2009}{\natexlab{a}})}\BibitemShut {NoStop}%
\bibitem [{\citenamefont {Sych}\ and\ \citenamefont
  {Leuchs}(2009{\natexlab{b}})}]{Sych:2009kl}%
  \BibitemOpen
  \bibfield  {author} {\bibinfo {author} {\bibfnamefont {D.}~\bibnamefont
  {Sych}}\ and\ \bibinfo {author} {\bibfnamefont {G.}~\bibnamefont {Leuchs}},\
  }\href@noop {} {\bibfield  {journal} {\bibinfo  {journal} {New J. Phys.}\
  }\textbf {\bibinfo {volume} {11}},\ \bibinfo {pages} {013006} (\bibinfo
  {year} {2009}{\natexlab{b}})}\BibitemShut {NoStop}%
\bibitem [{\citenamefont {Wittmann}\ \emph
  {et~al.}(2010{\natexlab{a}})\citenamefont {Wittmann}, \citenamefont
  {Andersen}, \citenamefont {Takeoka}, \citenamefont {Sych},\ and\
  \citenamefont {Leuchs}}]{Wittmann:2010mi}%
  \BibitemOpen
  \bibfield  {author} {\bibinfo {author} {\bibfnamefont {C.}~\bibnamefont
  {Wittmann}}, \bibinfo {author} {\bibfnamefont {U.~L.}\ \bibnamefont
  {Andersen}}, \bibinfo {author} {\bibfnamefont {M.}~\bibnamefont {Takeoka}},
  \bibinfo {author} {\bibfnamefont {D.}~\bibnamefont {Sych}}, \ and\ \bibinfo
  {author} {\bibfnamefont {G.}~\bibnamefont {Leuchs}},\ }\href@noop {}
  {\bibfield  {journal} {\bibinfo  {journal} {Phys. Rev. Lett.}\ }\textbf
  {\bibinfo {volume} {104}},\ \bibinfo {pages} {100505} (\bibinfo {year}
  {2010}{\natexlab{a}})}\BibitemShut {NoStop}%
\bibitem [{\citenamefont {Wittmann}\ \emph
  {et~al.}(2010{\natexlab{b}})\citenamefont {Wittmann}, \citenamefont
  {Andersen}, \citenamefont {Takeoka}, \citenamefont {Sych},\ and\
  \citenamefont {Leuchs}}]{Wittmann:2010pi}%
  \BibitemOpen
  \bibfield  {author} {\bibinfo {author} {\bibfnamefont {C.}~\bibnamefont
  {Wittmann}}, \bibinfo {author} {\bibfnamefont {U.~L.}\ \bibnamefont
  {Andersen}}, \bibinfo {author} {\bibfnamefont {M.}~\bibnamefont {Takeoka}},
  \bibinfo {author} {\bibfnamefont {D.}~\bibnamefont {Sych}}, \ and\ \bibinfo
  {author} {\bibfnamefont {G.}~\bibnamefont {Leuchs}},\ }\href@noop {}
  {\bibfield  {journal} {\bibinfo  {journal} {Phys. Rev. A}\ }\textbf {\bibinfo
  {volume} {81}},\ \bibinfo {pages} {062338} (\bibinfo {year}
  {2010}{\natexlab{b}})}\BibitemShut {NoStop}%
\bibitem [{\citenamefont {Ourjoumtsev}\ \emph {et~al.}(2006)\citenamefont
  {Ourjoumtsev}, \citenamefont {Tualle-Brouri}, \citenamefont {Laurat},\ and\
  \citenamefont {Grangier}}]{Ourjoumtsev:2006qa}%
  \BibitemOpen
  \bibfield  {author} {\bibinfo {author} {\bibfnamefont {A.}~\bibnamefont
  {Ourjoumtsev}}, \bibinfo {author} {\bibfnamefont {R.}~\bibnamefont
  {Tualle-Brouri}}, \bibinfo {author} {\bibfnamefont {J.}~\bibnamefont
  {Laurat}}, \ and\ \bibinfo {author} {\bibfnamefont {P.}~\bibnamefont
  {Grangier}},\ }\href@noop {} {\bibfield  {journal} {\bibinfo  {journal}
  {Science}\ }\textbf {\bibinfo {volume} {312}},\ \bibinfo {pages} {83}
  (\bibinfo {year} {2006})}\BibitemShut {NoStop}%
\end{thebibliography}

%

\end{document}